\documentclass[11 pt]{article}
\usepackage[T1]{fontenc}
\usepackage[utf8]{inputenc}
\usepackage{authblk}
\usepackage{graphicx}
\usepackage{dcolumn}
\usepackage{bm}
\usepackage{amssymb}
\usepackage{amsmath}
\usepackage{titlesec}
\usepackage{relsize}
\usepackage{mathtools}
\begin{document}
\title{Nonlinear Interference effect for Optical Clocks in Yb Atoms}
\author[1]{Tara Ahmadi\thanks{tara.ahmadi.smart@gmail.com}}
\author[1]{Sergei.A Pulkin}
\affil[1]{Physics faculty, St.Petersburg State University,198904 Petergof,St.Petersburg,Russia}
%\author{Tara Ahmadi\thanks{tara.ahmadi.smart@gmail.com}\ ,Sergei.A Pulkin,\ T.H.Yoon}
%\affiliation{Physics faculty, St.Petersburg State University,198904 Petergof,St.Petersburg,Russia,Physics faculty,Korea University,136-713 Seoul,Korea}
\date{\today}
\maketitle
\begin{abstract}
In this paper, two-photon absorption resonance has been investigated . We studied this phenomenon in three-level system with common upper level, two lower long-life levels and two quasi-resonant laser fields with different frequency spectrum. It has been shown that, the absorption coefficient in the far distance from the line on the frequency of two-photon resonance has considerable amplitude and narrow width. Frequency difference between two lower levels is transition frequency for optical clocks. This resonance is much less sensitive to the effects of noise components and probing fields compared with electromagnetically induced transparency. Stark shift of the two-photon resonance can be offset and reduced by the changes in the probing field amplitude. We recommended to use an additional laser with narrow line to stabilize the frequency at optical clock to get a line in sub-Hertz range.
\end{abstract}
\section{Introduction}
Nonlinear interference effect (NIE)[1] occurs in the presence of strong driving and probing fields with arbitrary amplitude at the adjacent transition. Important feature of this phenomenon is the presence of nonlinear interference term in polarization, responsible for two-photon transition. Electromagnetically induction transparency (EIT) has been used in optical clock at frequency between $S_1^0$ and $P_3^0$ in neutral bosons like strontium $S^{88}$[2] and ytterbium Yb[3]. Analysis of solutions for the absorption coefficient [3] showed that value of  EIT and its width are weakly dependent on width of laser line(frequency fluctuations of driving and probing fields ) in their mutual phasing. That difference helps to stabilize the frequency of laser. So by using this frequency difference, laser with narrow radiation line can be stabilized. This process helps to get optical clock with ultra-high stability at frequency resonance of EIT.
In [4], it has been shown that at far distance of absorption line, near EIT frequency resonance, we got two-photon absorption resonance. This resonance has large amplitude like amplitude of absorption resonance at the center of line.
Aim of this work is theoretically analysis of signal and frequency response of three-level system on EIT. We studied these features in monochromatic and noisy fields to implement the optical standards of frequency on forbidden transition in traps of Yb. We showed that two-photon absorption resonance not only has considerable amplitude compared with EIT resonance, but also narrow width when there are significantly different probabilities of transitions from the upper common level to the lower long-life levels. Two-photon absorption resonance is much less sensitive to the amplitude and phase noises of driving and probing field compared with EIT resonance, using this resonance to implement standards of frequency seems attractive. What is more two-photon absorption resonance is shifted in frequency from EIT resonance by the amount of Stark shift. But Stark shift can be offset by changes in amplitude of probing field, also two-photon absorption resonance exactly is at frequency resonance when Rabi frequency is equal to frequencies of driving and probing fields. It means when frequency difference between lasers is equal to frequency of forbidden transition.
We note that in [5], this method has been studied based on the solutions of Schrödinger equation for wave function in excitation scheme by Raman pulses. In general words, narrow width, large amplitude, stability and possibility to offset Stark shift augur well for the use of using two-photon absorption resonance in optical clocks.
\section{Density matrix of three-level atoms in strong driving and probing fields on adjacent transitions}
We studied steady solutions of element of density matrix for monochromatic driving field at 2-3 transition and probing field at 1-2 transition [6].  Equations for elements of density matrix in $\Lambda$-scheme for three-level system with common level 2 and long-life levels 1 and 3 by using rotating wave approximation (RWA) are:
\begin{equation}
\begin{split}
\dot{\rho}_{11} & = 2\left(\rho^{'}_{12}V^{''}_{12}-\rho^{''}_{12}V^{'}_{12}\right)+\gamma_{21}\rho_{22}+\gamma_{31}\rho_{33}\\
\dot{\rho}_{22} & = -2\left(\rho^{'}_{12}V^{''}_{12}-\rho^{''}_{12}V^{'}_{12}\right)-2\left(\rho^{''}_{23}V^{'}_{23}-\rho^{'}_{23}V^{''}_{23}\right)-\gamma_{21}\rho_{22}-\gamma_{23}\rho_{33}\\
\dot{\rho}_{33} & = 2\left(\rho^{''}_{23}V^{'}_{23}-\rho^{'}_{23}V^{''}_{23}\right)+\gamma_{23}\rho_{22}-\gamma_{31}\rho_{33}\\
\dot{\rho}^{''}_{12} & = -V^{'}_{12}(\rho_{22}-\rho_{11})+\rho^{'}_{13}V^{'}_{23}+\rho^{''}_{13}V^{''}_{23}-\Delta_{12}\rho^{'}_{12}-\Gamma_{12}\rho^{''}_{12}\\
\dot{\rho}^{'}_{12} & = -V^{''}_{12}(\rho_{22}-\rho_{11})+\rho^{'}_{13}V^{''}_{23}+\rho^{''}_{13}V^{'}_{23}+\Delta_{12}\rho^{''}_{12}-\Gamma_{12}\rho^{'}_{12}\\
\dot{\rho}^{''}_{23} & = -V^{'}_{23}(\rho_{33}-\rho_{22})-\rho^{'}_{13}V^{'}_{12}-\rho^{''}_{13}V^{''}_{12}+\Delta_{23}\rho^{'}_{23}-\Gamma_{23}\rho^{''}_{23}\\
\dot{\rho}^{'}_{23} & = V^{''}_{23}(\rho_{33}-\rho_{22})+\rho^{''}_{13}V^{'}_{12}-\rho^{'}_{13}V^{''}_{12}-\Delta_{23}\rho^{''}_{23}-\Gamma_{23}\rho^{'}_{23}\\
\dot{\rho}^{''}_{13} & = -\rho^{'}_{23}V^{'}_{12}+\rho^{''}_{23}V^{''}_{12}+\rho^{'}_{12}V^{'}_{23}-\rho^{''}_{12}V^{''}_{23}+(\Delta_{23}-\Delta_{12})\rho^{'}_{13}-\Gamma_{13}\rho^{''}_{13}\\
\dot{\rho}^{'}_{13} & = \rho^{'}_{23}V^{''}_{12}+\rho^{''}_{23}V^{'}_{12}-\rho^{'}_{12}V^{''}_{23}-\rho^{''}_{12}V^{'}_{23}-(\Delta_{23}-\Delta_{12})\rho^{''}_{13}-\Gamma_{13}\rho^{'}_{13}\\
\end{split}
\end{equation}
Where $\Delta$=difference between frequency of laser and transition frequency $\omega$. $\gamma$ – transition probability, $\Gamma$-line width, elements with a bar means real part and two bars means imaginary part. Matrix elements of interaction energy which describe interaction of fields with atoms during k$\longrightarrow$ i transition are: $ V^{'}_{12}(t)=Re(V_1(t)) \ , \ V^{''}_{12}(t)=Im(V_1(t)) \ ,\ V^{'}_{23}(t)=Re(V_2(t)) \ , \ V^{''}_{23}(t)=Im(V_2(t))$.
Where V(t) = slow term of matrix elements of interaction energy, introduced for monochromatic fields of Rabi frequency. For general situation, with different phases, phase can be engaged in equations:
\begin{equation}
V_1(t)=V_{10}\exp(-i\varphi_{1})\ , \ V_2(t)=V_{20}\exp(-i\varphi_{2})
\end{equation}
where\ $V_{10}=\frac{\mu_{12}E_{i}}{\hbar}$ amplitude,\ $\mu_{12}$-matrix element of dipole transition,\ $E_{i}$-amplitude electrical field and\ $\varphi_{i}$-phase of laser field.
In case of closed atomic structure$\Sigma\rho_{ii}=1$.
For steady state, when the derivatives of the matrix elements of the time-zero, the system of equations becomes algebraic system with constant or slightly varying coefficients of the time.
Condition for the steady state has the form [6]
\begin{equation}
\max\left|\frac{\frac{\partial V_{ij}}{\partial t}}{V_{ij}}\right|\ll\min|\Gamma_{ij}+i\Delta_{ij}|
\end{equation}
In the case of long-life lower levels with line width\ $\Gamma_{31}=0$ and\ $\Delta_{31}=\Delta_{12}-\Delta_{32}=0$, steady solution is acceptable only for monochromatic field not for noises of laser field.it means that every fluctuation in the frequencies of the fields or their difference led to perturbation of steady state. Also in this case amplitude and width of EIT resonance depend on width of laser line deeply, which is obvious from the analysis of solutions.
Steady solutions for matrix equation(1)are:
\begin{equation}
\begin{split}
\rho_{31} & = \left(\frac{V_{32}V_{12}}{\Gamma^{'}_{13}}+Q_{31}\right)\left(\frac{n_{12}}{\Gamma^{'}_{12}}+\frac{n_{32}}{\Gamma^{'}_{32}}\right)\\
\rho_{21} & = \left(\frac{V_{21}n_{12}}{\Gamma^{'}_{12}}+g_{21}\right)+\left(\frac{V_{21}|V_{32}|^{2}n_{32}}{(\Gamma^{'}_{13}+Q_{31})\Gamma^{'}_{12}\Gamma^{'}_{32}}\right)\\
\rho_{32} & = \left(\frac{V_{32}n_{32}}{\Gamma^{'}_{32}}+g_{32}\right)+\left(\frac{V_{32}|V_{21}|^{2}n_{12}}{(\Gamma^{'}_{13}+Q_{31})\Gamma^{'}_{12}\Gamma^{'}_{32}}\right)\\
\end{split}
\end{equation}
where\ $\Gamma^{'}_{12}=\Gamma_{12}+i\Delta_{12},\ \Gamma^{'}_{32}=\Gamma_{32}+i\Delta_{32},\ \Gamma^{'}_{13}=\Gamma_{13}+i(\Delta_{32}-\Delta_{12})$ and;
\begin{equation}\nonumber
\begin{split}
Q_{31} & =\frac{|V_{21}|^{2}}{\Gamma^{'}_{32}+\frac{|V_{32}|^{2}}{\Gamma^{'}_{12}}}\\
Q_{21} & =\frac{|V_{32}|^{2}}{\Gamma^{'}_{13}+\frac{|V_{21}|^{2}}{\Gamma^{'}_{12}}}\\
Q_{32} & =\frac{|V_{21}|^{2}}{\Gamma^{'}_{13}+\frac{|V_{32}|^{2}}{\Gamma^{'}_{12}}}\\
\end{split}
\end{equation}
Population difference\ $n_{12}=\rho_{11}-\rho_{22}, n_{32}=\rho_{33}-\rho{22}$ can be find as:
\begin{equation}
\begin{split}
n_{12} & =\frac{(-A_{22}f_{3}+A_{12}f_{6})}{(A_{11}A_{22}-A_{21}A_{12})}\\
n_{32} & =\frac{(-A_{11}f_{6}+A_{21}f_{3})}{A_{11}A_{22}-A_{21}A_{12}},\\
\end{split}
\end{equation}
where$A_{11}=-f_{1}-2a_{21}+b_{12}\ ,A_{12}=f_{2}+a_{23}-2b_{13}\ ,A_{22}=-f_{4}-2a_{23}+b_{13}$.
f depends on transition population between levels(d) and non-coherent pumping:
\begin{equation}
\begin{split}
f_{1} & = \frac{(4d_{12}+2d_{21}+2d_{13}+d_{31}+d_{23}-d_{32})}{3}\\
f_{2} & = \frac{(-2d_{12}+2d_{21}-d_{13}-2d_{31}+d_{23}+2d_{32})}{3}\\
f_{3} & = \frac{(-2d_{12}+2d_{21}-d_{13}+d_{31}+d_{23}-d_{32})}{3}\\
f_{4} & = \frac{(-d_{12}+d_{21}+d_{13}+2d_{31}+2d_{23}+4d_{32})}{3}\\
f_{5} & = \frac{(2d_{12}+d_{21}-2d_{13}-d_{31}+2d_{23}-2d_{32})}{3}\\
f_{6} & = \frac{(d_{12}-d_{21}-d_{13}+d_{31}-2d_{23}+2d_{32})}{3}\\
\end{split}
\end{equation}
Populations of one-photon and two-photon absorptions are:
\begin{equation}
\begin{split}
a_{ij} & = 2Re\left\{\frac{|V_{ij}|^2}{\Gamma^{'}_{ij}+g_{ij}}\right\}\\
b_{13} & = 2Re\left\{\frac{|V_{12}V_{23}|^2}{(\Gamma^{'}_{13}+Q_{31})\Gamma^{'}_{12}\Gamma^{'}_{32}}\right\}\\
\end{split}
\end{equation}
For $\Lambda$-system,populations\ $f_{ik}$ are:
\begin{equation}
\begin{split}
f_{1} & = \frac{(4d_{12}+2\gamma_{21}+2d_{13}+\gamma_{31}+\gamma_{23})}{3}\\
f_{2} & = \frac{(-2d_{12}+2\gamma_{21}-d_{13}-2\gamma_{31}+\gamma_{23})}{3}\\
f_{3} & = \frac{(-2d_{12}+2\gamma_{21}-d_{13}+\gamma_{31}+\gamma_{23})}{3}\\
f_{4} & = \frac{(d_{12}+\gamma_{21}+d_{13}+2\gamma_{31}+2\gamma_{23})}{3}\\
f_{5} & = \frac{(2d_{12}+\gamma_{21}-2d_{13}-\gamma_{31}+2\gamma_{23})}{3}\\
f_{6} & = \frac{(d_{12}-\gamma_{21}-d_{13}+\gamma_{31}-2\gamma_{23})}{3}\\
\end{split}
\end{equation}
We compare the solutions obtained by direct integration of the time, followed by Fourier transformation, with analytical solutions for the case of monochromatic fields. Also we investigated the numerical solutions for the time variation of the density matrix elements for both monochromatic and for noisy fields. Polarization spectrum is detected from the time course of the diagonal elements using the Fourier transform(CFFT).
\begin{equation}
P(\omega)=CFFT\left(\rho^{'}_{ik}(t)-i\rho^{''}_{ik}(t)\right)
\end{equation}
Max load capacity spectrum has the form:
\begin{equation}
I(\omega)=Re(P(\omega))^{2}+Im(P(\omega))^2
\end{equation}
Absorption coefficient for probing field during 1$\longrightarrow$2 transitions is obtained from relationship of the detuning of the spectral components from frequency of probing field:
\begin{equation}
k_{\upsilon}=\rho^{''}_{12}(\Delta_{12})N_{0}\frac{\lambda^2}{4\pi}\frac{1}{V_{10}}
\end{equation}
where$N_{0}$-concentration of normal atoms and,\ $\lambda$-wave length.
For the weak probing field $V_{10}\ll V_{20}$, in the absence of incoherent pumping between levels 2 and 3 for absorption is a relatively simple expression, gotten from general answer of system (4)[6]:
\begin{equation}
W_{12}(\omega)=2\hbar\omega Re\left\{\frac{\Gamma_{31}+i(\Delta_{12}-\Delta_{23})}{(\Gamma_{21}+i\Delta_{12})(\left(\Gamma_{31}+i(\Delta_{12}-\Delta_{23})\right)+|V_{20}|^2}\right\}
\end{equation}
Typical absorption section around two-photon absorption resonance has been shown in Figure.2 for these parameters of fields and atom:$ \gamma_{12}=1; \gamma_{23}=0.01; \gamma_{31}=0;V_{10}=0.0001;V_{20}=0.1;\Delta_{32}=5\gamma_{12}$.
As can be seen, for large \ $\Delta_{32}=5\gamma_{12}$ in the far distance of absorption line, is narrow two-photon absorption resonance. This resonance shifted relatively EIT resonance in the amount of Stark’s shift. Stark’s shift is equal to\ $ \frac{(V_{20})^2}{\Delta_{12}}$. Calculations showed that amplitude of two-photon absorption resonance is equal to the amount of absorption coefficient at the center of line for one-photon absorption. [4]
In this work unlike [4], we studied when transition possibilities are very different $\gamma_{21}\gg\gamma_{23}$. We proved that amplitude of this resonance is large and its width is narrow.
\section{Absorption spectrum in the excitation of the atomic system for noisy and monochromatic fields}
Time dependences of density matrix elements give us information about dynamics of processes responsible for two-photon absorption, formation processes of dark EIT resonances and coherent trapping (CPT). In this work, we also studied dynamics of these processes in monochromatic and noisy fields also the effects of non-monochromatic fields on absorption spectrum. These time dependences show that matrix elements responsible for atom’s polarization in weak probing field oscillate with frequency of\ $\delta=\delta_{0}-\left(\frac{(V_{20})^2}{\Delta_{12}}\right)$ where\ $\delta_{0}=\Delta_{12}-\Delta_{23}$.
Coherency $(\rho(t))$ oscillates with this frequency and makes $\rho_{12}(t)$ to oscillate with this frequency too but in anti-phase direction. Oscillation decays with constant coefficient$ \frac{(V_{20})^2}{\Delta_{12}}$.\\
For the case of non-chromatic fields, we added amplitude and phase of noise to the equations of fields.
\begin{equation}
\begin{split}
V_{1}(t)& = V_{10}(1+0.0001)\sum_{n=0}^{100}\left[\exp(-i(-50+n)S_{1}t+\varphi_{n})\right]\\
V_{2}(t)& = V_{20}(1+0.0001)\sum_{m=0}^{100}\left[\exp(-i(-50+m)S_{2}t+\varphi_{m})\right]\\
\end{split}
\end{equation}
Where $S_{i}$-frequency difference between monochromatic components,$\varphi_{n},\varphi_{m}$ phases.
Polarization time dependence has been shown in Figure 2.:a)$\Delta_{12}=5.004 \ (\delta=0)$-position of two-photon resonance and b)$\Delta_{12}=5.000 \ (\delta=\frac{(V_{20})^2}{\Delta_{12}})$-position of resonance for monochromatic and noisy fields.
Calculations showed that amplitude of two-photon absorption resonance is larger than EIT amplitude. Also for this level noises are offset while two-photon resonance is valid. Considering the effects of noises on the form of resonance in absorption spectrum of probing field. In Figure.3 you can see absorption spectrum with these parameters:$\gamma_{12}=1;\gamma_{23}=0.01;\gamma_{31}=0;V_{10}=0.0001;V_{20}=0.141421;S_{1}=S_{2}=0.0005;m=n;\Delta_{32}=5\gamma_{12}$.
In this figure also has been shown the absorption spectrum of monochromatic fields (dot line).
You can see in the Figure.3 that amplitude and width of noise of two-photon absorption resonance stored. In this time amplitude of EIT resonance offsets or strongly decays. In other hands, amplitudes obtained numerical and analytical are the same when timescale is large.
\section{Atomic clocks in cold atoms of ytterbium}
Probing field is quasi-resonant with 1-2 transition (for the case of Yb atoms –this resonance occurs between $S_{0}^1$ and $P_{0}^1$). Divining field is quasi-resonant with 1-3 transition (for the case of Yb atoms –this resonance occurs between $P_{0}^3$ and $P_{0}^1$). 1-3 is a forbidden transition (forbidden for bosons, life-time for fermions is several minutes and for bosons between $S_0^1$-$P_0^3$ life-time is several years).Absorption spectrum has been shown in Figure.4 for the parameters:$\gamma_{12}=1;\gamma_{23}=10^{-8};\gamma_{31}=0;V_{10}=10^{-3}V_{20}$(dot line) and $V_{10}=V_{20}$(dashed line).
Probability of transition between $S_0^1-P_1^1$ in Yb atoms is equal with$\gamma_{12}=2\times10^{8}s^{-1}$; for $P_1^1-P_0^3$ transition,magnetic matrix is obvious $[3]:<2|M_0|3>=0.022\mu_{B\mu}$.In Figure.4 we have $\gamma_{23}=10^{-8};\gamma_{12}=2Hz$.\\
As you can see in Figure.4.a) in the far distance of absorption line, narrow two-photon resonance occurs which its amplitude is equal with the amplitude of absorption coefficient at the center of one-photon absorption. This resonance has been shifted from the EIT resonance by the amount of Stark’s shift. The width of resonance and Stark’s shift in this case are equal to $ \frac{(V_{20})^2}{\Delta_{12}}\sim 10^{-6}HZ$.\\
Stark’s shift can be offset by changing the amplitude of probing field. When the Rabi frequencies of probing and driving field are the same, two-photon resonance is exactly on the zero difference between $\Delta_{32}-\Delta_{32}=0$. For the range of mille-Hertz, intensity of driving field $\sim4 mW/cm^2$ and the amount of field term (Rabi frequency) is $V_{20}=10^3 s^{-1}$.
\section{Result}
Nonlinear two-photon interference resonance has been studied numerically and analytically based on characters of optical clocks in cold atoms of Ytterbium.
Time and frequency dependencies of matrix elements show that two-photon absorption resonance is less sensitive than EIT resonance to the noises of the fields. Also we showed that Stark's shift can be offset by changing amplitude of probing field.Complete offset occurs when Rabi frequencies of probing and driving field are the same.

\begin{figure}[b]
\begin{minipage}[h]{1\linewidth}
{\includegraphics[width=1\linewidth]{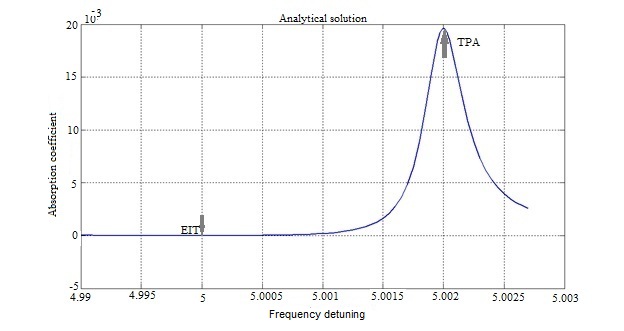}}{a}
{\includegraphics[width=1\linewidth]{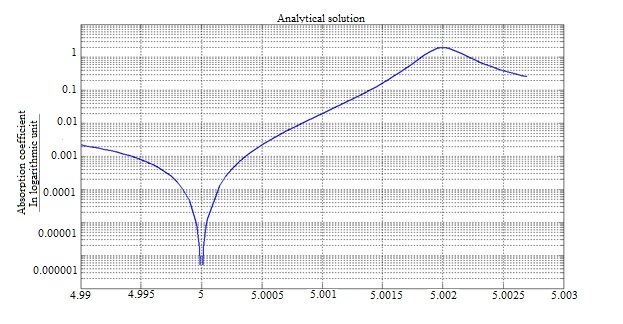}}{b}\\
\caption{ Absorption coefficient from frequency detuning}
\end{minipage}
\end{figure}
\begin{figure}[b]
\begin{minipage}[b]{1\linewidth}
{\includegraphics[width=0.5\linewidth]{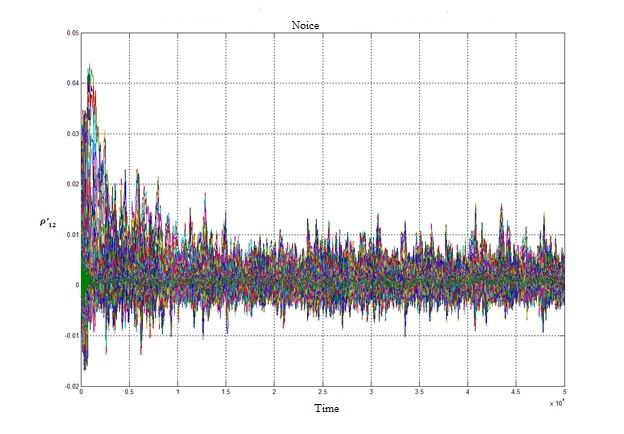}}{a}
{\includegraphics[width=0.5\linewidth]{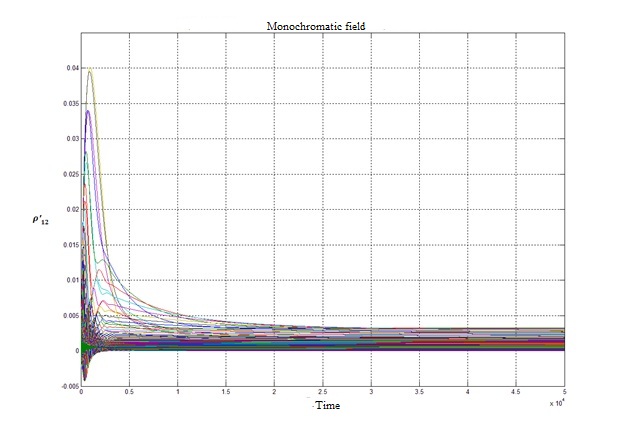}}{b}
{\includegraphics[width=0.6\linewidth]{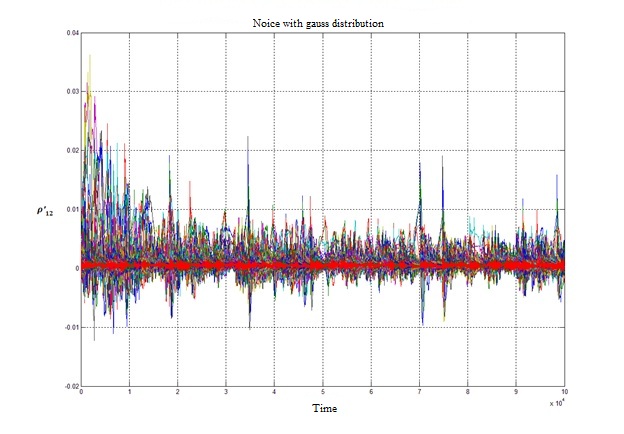}}{c}
\caption{ Polarization of resonance signal from time}
\end{minipage}
\end{figure}
 \begin{figure}[b]
\begin{minipage}[h]{1\linewidth}
\center{\includegraphics[width=0.7\linewidth]{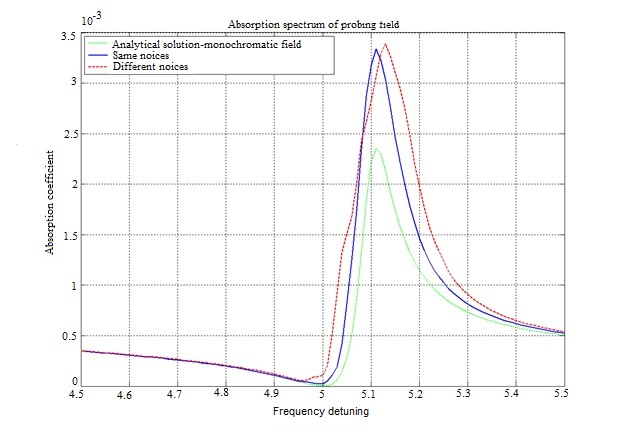}}\\
\caption{ Absorption coefficient from frequency detuning}
\end{minipage}
\end{figure}
\begin{figure}[b]
\begin{minipage}[h]{1\linewidth}
\center{\includegraphics[width=0.7\linewidth]{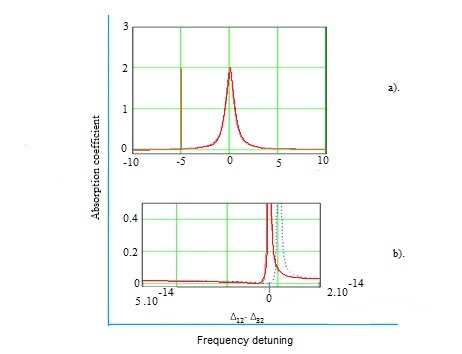}}\\
\caption{ Absorption coefficient for Ytterbium}
\end{minipage}
\end{figure}
\end{document}